\documentclass[runningheads]{svmult}

\usepackage{makeidx}   
\usepackage{graphicx}  
\usepackage{subeqnar}  
\usepackage{multicol}  
\usepackage{physprbb}  
\makeindex             

\def\ltsima{$\; \buildrel < \over \sim \;$}
\def\lsim{\lower.5ex\hbox{\ltsima}}
\def\gtsima{$\; \buildrel > \over \sim \;$}
\def\gsim{\lower.5ex\hbox{\gtsima}}

\begin{document}

\title*{Iron line emission in X--ray afterglows}
\titlerunning{Iron line emission in X--ray afterglows}
\author{Davide Lazzati\inst{1,2} \and Gabriele Ghisellini\inst{1} \and
Mario Vietri\inst{3} \and Fabrizio Fiore\inst{4} \and
Luigi Stella\inst{4}}
\authorrunning{D. Lazzati et al.}
\institute{Osservatorio Astronomico di Brera, Via Bianchi 46
I--23807 Merate (Lc), Italy \and Present Address: Institute of 
Astronomy, Cambridge CB3 0HA, UK
\and Universit\`a di Roma 3, via della Vasca Navale 84, I--00147 Roma, 
Italy \and Osservatorio Astr. di Roma, via Frascati 33, I--00040
Monteporzio Catone, Italy}

\maketitle

\begin{abstract}
Recent observations of X--ray afterglows reveal the presence of 
a redshifted K$\alpha$ iron line in emission in four bursts.
In GRB 991216, the line was detected by the low
energy grating of {\it Chandra}, which showed the line to be broad, with
a full width of $\sim$15,000 km s$^{-1}$.
These observations indicate the presence of a $>1$ $M_\odot$ 
of iron rich material in the close vicinity of the burst, most likely
a supernova remnant.
The fact that such strong lines are observed less than a day after the 
trigger strongly limits the size of the remnant, which must be very compact. 
If the remnant had the observed velocity since the supernova
explosion, its age would be less than a month.
In this case nickel and cobalt have 
not yet decayed into iron. We show how to solve this paradox.
\end{abstract}

\section{Introduction}

There are now four bursts displaying evidence
of an emission line feature during the X--ray afterglow:
GRB 970508  (Piro et al., 1999);
GRB 970828 (Yoshida et al., 1999); 
GRB 991216 (Piro et al., 2000, hereafter P2000); 
GRB 000214 (Antonelli et al., 2000). 
These lines have been observed 8--40 hours after the burst explosion,
have a large equivalenth width (0.5--2 keV) and a flux of about
$10^{-13}$ erg cm$^{-2}$ s$^{-1}$.
Given these properties, each iron atom has to produce at least 2000 
line photons, in order not to exceed 0.1 $M_\odot$ mass of emitting iron,.
Fast recombination and ionization is therefore required.
The line of GRB991216 is resolved in the $Chandra$ gratings, 
with a width $0.05c$ (P2000). 
As discussed by Lazzati et al. (1999), the detection of
the line implies the presence of a sizable fraction of a solar mass of 
iron concentrated in the vicinity of the GRB site. 
This is naturally accounted for in the SupraNova scenario 
(Vietri \& Stella 1998).

\section{General Constraints}
\noindent {\bf The size problem} 
~~~~If the line is detected after $t_{\rm obs}$ from the burst,
the line emitting material must be located within a distance $R$ given by:
\begin{equation}
R\, \le\, {ct_{\rm obs} \over 1+z} \, {1\over 1-\cos\theta}\, \simeq \,
{ 1.1\times 10^{15}\over 1+z}  \, {t_{\rm obs} \over10\, {\rm h} } \,\, 
{1\over 1-\cos\theta}\,\, {\rm cm},
\label{eq:rag}
\end{equation}
where $\theta$ is the angle between the line emitting 
material and the line of sight at the GRB site.
This limit implies a large scattering optical depth:
\begin{equation}
\tau_{\rm T}\, = \, {\sigma_{\rm T} M \over 4\pi R^2 \mu m_{\rm p}}
\ge \, 54 \, { (M/M_\odot)(1+z)^2(1-\cos\theta)^2 \over \mu \, 
(t_{\rm obs}/10\, {\rm h})^2},
\end{equation}
where $\mu$, is the mean atomic weight of the material.

\noindent{\bf The kinematic problem}
~~~~For a radial velocity of the remnant of $v=10^9v_9$ cm s$^{-1}$
the time elapsed from the supernova (SN) is $t_{\rm SN} \simeq 
12.5 (t_{\rm obs}/10{\rm hr}) / [(1+z)(1-\cos\theta)v_9]$ days.
Such short times implies that most of the 
$^{56}$Co nuclei (and a fraction of the $^{56}$Ni nuclei) have not yet decayed 
to $^{56}$Fe (half--life of 77.3 and 6.08 days, respectively, see Vietri et 
al. 2000).

\noindent{\bf Line emission rate}
~~~~We can derive the photon line luminosity
by estimating the volume $V_{\rm em}$ effectively contributing to 
the line emission, and assuming a given iron mass.
If the layer contributing to the emission has $\tau_{\rm T}\sim 1$
(to avoid Compton broadening), and in this layer $\tau_{\rm FeXXVI}\sim$a
few (to efficiently absorb the continuum),
we have $V_{\rm em}=S/(\sigma_{\rm T} n_{\rm e})$, where $S$ is the 
emitting surface. The line emission rate from $V_{\rm em}$ is then:
\begin{equation}
\label{rate}
\dot N_{\rm Fe} = {N_{\rm Fe}\over t_{\rm rec}} =
{Sn_{\rm Fe} \over 1.3\times 10^{11} T_7^{3/4} \sigma_{\rm T} } \sim 
3\times 10^{53} {(M_{\rm Fe}/M_\odot) \over T_7^{3/4} \Delta R_{15}}
\quad {\rm s^{-1}},
\label{eq:ndot}
\end{equation}
where the {\it total} volume is $V=S \Delta R$ (slab or shell geometry). 

\noindent{\bf Mass}
~~~~Eq.~\ref{eq:ndot} shows that the total iron mass must be a sizable 
fraction of a solar mass in order to give rise to the observed line 
photon luminosity of $4\times 10^{52}$~s$^{-1}$.
Notice also that Eq.~\ref{eq:ndot} establishes 
that the line emitting material must be a SNR: no other known astrophysical
object contains this iron mass.

\section{Models}
\noindent{\bf The wide funnel}
~~~~Consider a wide funnel excavated in a young plerionic remnant.
This solves the {\it size problem}, since  it extends to large radii 
but can maintain the time--delay
contained because it is built close to the polar axis (see Fig.~2).
Fixing the line photon rate (Eq. 3) yields $R = 6\times 10^{15}$ cm, 
and thus an opening angle $\theta = 48^\circ$ to fit the time--delay.
Assuming a cone geometry for simplicity, we can rewrite 
Eq.~\ref{eq:ndot} as:
\begin{equation}
\dot N_{\rm Fe} \, =\, 3.3\times 10^{52} 
{(M_{\rm Fe}/M_\odot) \over T_7^{3/4}  (R_{15}/6)} \tan{\theta}
\quad {\rm s^{-1}}.
\label{eq:ndot2}
\end{equation}
This is a lower limit, since a parabolic funnel has a larger 
surface and we neglected the (likely) density stratification inside 
the remnant. 
Consider now the kinematic properties of the funnel. 
We expect radiation pressure to exert a force parallel 
to the surface accelerating the layer with $\tau_{\rm T} = 1$.
The absorbed fluence 
$E_{\rm ion}$ accelerates the funnel layer to 
$v_{\rm f} = (2 E_{\rm ion}/M_{\rm layer})^{1/2} \sin\phi \simeq  
10^4 E_{\rm ion, 50}^{1/2}\sin\phi$~km~s$^{-1}$ 
if $R=6\times 10^{15}$~cm. $\phi$ is the angle between the funnel's normal and 
the incoming photons. Thus, we expect ablation by radiation pressure 
to be able to propel the reflecting layer to velocities comparable to 
those seen in GRB991216.

\noindent{\bf Back illuminated equatorial material}
~~~~The model above assumes that a SN explosion
preceded the GRB by some months.
We now explore the possibility of a simultaneous GRB--SN
explosion. Assume that a GRB ejects and accelerates 
a small amount of matter in a collimated cone, while a large amount of 
matter is instead ejected, at sub--relativistic speeds,
along the progenitor's equator.
Massive star progenitors are inevitably surrounded
by dense material produced by strong winds of mass loss rates 
$\dot m_{\rm w}=10^{-5} \dot m_{\rm w, -5}$
and velocity $v_{\rm w}=10^7v_{\rm w,7}$.
This wind scatters back a fraction of the photons produced by the bursts
and its afterglow (Thompson \& Madau 2000). The scattered luminosity 
$L_{\rm scatt}$ is constant, since there is an equal number 
of electrons in a shell of constant width $\Delta R$
(for a density profile $\propto R^{-2}$).
This luminosity is of order:
\begin{equation} 
L_{\rm scatt} \, \sim \, m_{\rm p} c^2\,  
{\dot m_{\rm w}\over m_{\rm p} v_{\rm w}/c}\, =\, 
1.8\times 10^{45}\,
{\dot m_{\rm w, -5}\over v_{\rm w,7}} \,\,\,\, {\rm erg\,\, s^{-1}}.
\end{equation}
Scattered photons illuminate the expanding 
equatorial matter after a time $2R/c$, giving rise to the line emission.
Since in this case the SN and GRB explosions are supposed to be
simultaneous, the emitting iron must be produced directly by the 
SN and not through the nickel decay. Iron ($^{54}$Fe) is
directly synthesized for high neutronization of the material at the 
SN shock.

\section{Conclusions}
The recently detected features in the X--ray afterglow of GRBs impose
strong constraints on models, the most severe being how to arrange a
large amount of iron close to the GRB site, while
avoiding at the same time a large Thomson scattering opacity.
This limit applies to all bursts showing a line feature.
An additional limit comes from the $Chandra$ observation of a $broad$ line
in GRB 991216.
These observations require a very large amount of iron, 
known to be contained only in SNe. 
We have described two models. The ``wide funnel" model is in better 
agreement with observations: its geometry solves the size problem, and the 
acceleration of the line emitting material by grazing incident 
photons solves the kinematic problem, allowing the remnant to be a few 
months old (enough for most cobalt to have decayed into iron).
This model implies that the GRB progenitors are massive stars 
exploded as SNe some months before the burst, inundating
the surroundings of the burst with iron rich material.
This two--step process and the time--delay between the two steps are 
exactly what is predicted in the SupraNova 
scenario of Vietri \& Stella (1998).


\begin{thebibliography}{8.}
\addcontentsline{toc}{section}{References}
\bibitem{} Antonelli A. et al., 2000, ApJ, 545, L39
\bibitem{} Lazzati D., Campana S. \& Ghisellini G., 1999, MNRAS, 304, L31
\bibitem{} Piro L., et al., 1999, ApJ, 514, L73
\bibitem{} Piro L., et al., 2000, Science, 290, 955
\bibitem{} Thompson C. \& Madau P., 2000, ApJ, 538, 105
\bibitem{} Vietri M. \& Stella L., 1998, Ap.J.Lett., 507, L45
\bibitem{} Vietri M., Ghisellini G., Lazzati D., Fiore F. \& Stella L.,
	2001, ApJL, in press (astro-ph/0011580)
\bibitem{} Yoshida, A., et al., 1999, Astr. Ap. Suppl., 138, 433
\end{thebibliography}
\end{document}